\documentclass[3p,twocolumn]{elsarticle}

\usepackage{graphicx}

\usepackage{amssymb}

\journal{ARENA Proceedings}

\begin{document}

\begin{frontmatter}



\title{Air Shower Measurements with LOFAR}


\author[RU]{A.~Horneffer\corref{AH}}  
  \cortext[AH]{Corresponding author.}\ead{A.Horneffer@astro.ru.nl}
\author[RU]{L.~B\"ahren}
\author[RU]{S.~Buitink}
\author[RU,ASTRON]{H.~Falcke}
\author[RU]{J.R.~H\"orandel}
\author[RU]{J.~Kuijpers}
\author[RU]{S.~Lafebre}
\author[RU]{A.~Nigl}
\author[KVI]{O.~Scholten}
\author[RU,KVI]{K.~Singh}

\address[RU]{Department of Astrophysics/IMAPP, Radboud University Nijmegen, 6500 GL Nijmegen, The~Netherlands}
\address[ASTRON]{ASTRON, 7990 AA Dwingeloo, The Netherlands}
\address[KVI]{ Kernfysisch Versneller Instituut, NL-9747 AA Groningen, The Netherlands}

\begin{abstract}
Air showers from cosmic rays emit short, intense radio pulses.
LOFAR is a new radio telescope, that is being built in the Netherlands and Europe.
Designed primarily as a radio interferometer, the core of LOFAR will have
a high density of radio antennas, which will be extremely well calibrated.
This makes LOFAR a unique tool for the study of the radio properties of
single air showers. 

Triggering on the radio emission from air showers means detecting a short
radio pulse and discriminating real events from radio interference. 
At LOFAR we plan to search for pulses in the digital data stream - either from 
single antennas or from already beam-formed data - and calculate several 
parameters characterizing the pulse shape to pick out real events in a second stage.
In addition, we will have a small scintillator array to test and confirm the
performance of the radio only trigger.

\end{abstract}


\end{frontmatter}



\section{Introduction}

It has been known since 1965 that cosmic ray air showers emit short radio 
pulses\cite{Jelley}.
LOPES, a {\bf LO}FAR {\bf P}rototyp{\bf e} {\bf S}tation, was the first experiment 
that has proven that with fast ADCs and modern computer technology it is possible 
to measure these radio pulses even in the presence of relatively strong 
RFI\footnote{{\bf r}adio {\bf f}requency {\bf i}nterference}\,\cite{nature}.

\begin{figure}
\begin{center}
\includegraphics[width=0.95\linewidth]{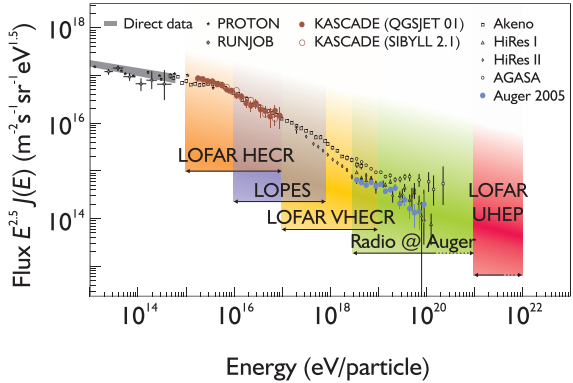}
\caption{
 \label{fig:crflux}
Cosmic ray energy spectrum with the measurement modes of LOFAR. In the HECR mode we 
look for air showers in beam-formed data, in VHECR mode we look for air showers in
single channel data and in UHEP mode we look for radio pulses from particles 
hitting the moon. LOPES and Auger are related experiments.
   }
\end{center}
\end{figure}

LOFAR, the {\bf Lo}w {\bf F}requency {\bf Ar}ray, is a new radio telescope for the
frequency range of 10\,MHz to 270\,MHz, that is being built in the Netherlands and Europe.
Designed as a radio interferometer it will consist of more than 40 stations with 
fields of comparatively simple antennas. These stations  are placed
densely in the core of LOFAR and at increasing distances the further away from the
core a station is. The core will have at least 18 stations of 48/96 dual/single 
polarization antennas in a roughly 2\,km by 3\,km area. 
To enable the radio astronomy application LOFAR will have a very precise calibration.
While this setup would not be cost-effective for pure air shower measurements it gives 
us an unique possibility to study the radio emission from single air showers in 
great detail.
Figure~\ref{fig:crflux} shows the different energy ranges in which LOFAR will
measure cosmic rays. The HECR\footnote{high energy cosmic rays} and the 
VHECR\footnote{very high energy cosmic rays} mode both measure air showers, they 
differ only in the method of triggering, but share the same post-processing and
analysis software. 

\section{LOFAR Air Shower Trigger}

Air shower radio pulses are short ($<$20\,ns) pulses, that are beamed in the 
propagation direction of the air shower and illuminate an area comparable to the size 
of the particle disc. In addition the radio pulses do not arrive on the ground in a 
plane but with some curvature. Compared to a plane wave the radio pulses arrive later
with increasing distance from the shower axis.
These properties can be used to distinguish air shower pulses from pulsed RFI. 

In LOFAR there are to paths for the data. The signals from each dipole are 
digitized and then are both sent to the so called transient buffer boards (TBB) for
storage in a memory ring-buffer and are processed in a continuous data stream. In 
this stream the data is converted into frequency bands, these bands are then
beam-formed in a given direction at the station level and the beam-formed data from
each station is sent to a {\bf ce}ntral {\bf p}rocessing facility (CEP). 

In the HECR mode we look at the beam-formed data from single stations. At CEP the 
beam-formed data is converted back into time-domain for full time resolution and 
searched for air shower candidate pulses. When a suitable candidate is found the 
data from the corresponding TBBs is saved to a file for off-line processing. As the 
beam-forming reduces the solid angle but increases the sensitivity, this mode is
useful for low energy showers.

At higher energies, in the VHECR mode, we look for pulses in the data streams of 
single channels. This is done in three levels: the first level runs on the hardware
of the TBBs. There the data from each channel is first filtered with up to three
so called infinite impulse response (IIR) filters, that can be configured as high-pass,
low-pass, or notch filters. The filtered data is then searched for pulses by evaluating 
the following equation:
\begin{equation}
|x_i| > \mu_i + a\sigma_i
\label{eq:first}
\end{equation}
(With $x_i$ the input data, $\mu_i$ and $\sigma_i$ the mean and standard deviation 
of $|x_i|$ , and $a$ a parameter.) As the input data is Gaussian distributed 
$\mu_i$ is proportional to $\sigma_i$, so this equation can be simplified to:
\begin{equation}
|x_i| > b\mu_i 
\label{eq:second}
\end{equation}
which greatly eases implementation on the TBBs. For every $x_i$ above threshold a 
counter is increased by 4 and decreased by one otherwise, unless it is zero. So
by testing if this counter is above a given value one can detect pulses in which several
values close to each other are above threshold. If such a pulse is detected the
pulse parameters {\em position in time, height, width, sum, average before} and 
{\em average after the pulse} are calculated and sent in a message to 
the computer controlling the station. On this computer the second trigger level runs. 
In this an incoming message is first checked if the pulse parameters disqualify
a pulse for an air shower candidate. With the remaining pulses a simple coincidence
check is done to detect air shower candidates. If the available computing power
permits we then can do a direction fit or look at the pulse height pattern on the 
ground to identify RFI pulses. If a good air shower candidate is identified all
TBBs of the station are notified and their data are written to disk. Information about
weak and good candidates is sent to CEP, which then can decide if a coincidence of weak
pulses in several stations indicates an air shower in between them, or if a strong 
air shower warrants saving the data from additional stations.

\section{Test Measurements}

\begin{figure*}
\begin{center}
\includegraphics[width=0.6\linewidth]{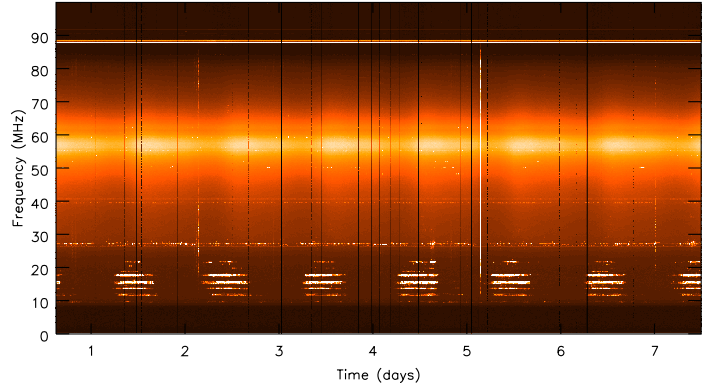}
\caption{
 \label{fig:dynspec}
Dynamic spectrum generated from LOFAR time-series data. The white horizontal line at 
88\,MHz is caused by a FM-transmitter, the white stripes between 8\,MHz and 20\,MHz 
by short-wave transmitters.
Black vertical lines are caused by problems in the prototype data acquisition software, 
white vertical lines (e.g. at 5.1 days) show times of intense interference.
The daily brightness variation is due to the galactic plane moving in and out of the
field of view.
   }
\end{center}
\end{figure*}

Since late 2007 there is already a test setup available, with one station of 48 
dual polarization antennas and three stations with 16 antennas each. Two stations 
are equipped with TBBs that can collect data from 16 antennas each, for a total
of 64 channels.
Figure \ref{fig:dynspec} shows a dynamic spectrum generated from LOFAR time-series 
data. Except for the short-wave band and one FM transmitter there is little RFI 
background for most of the time. Only at a few time intervals there was strong RFI 
present, that would make air shower measurements impossible.

\begin{figure}
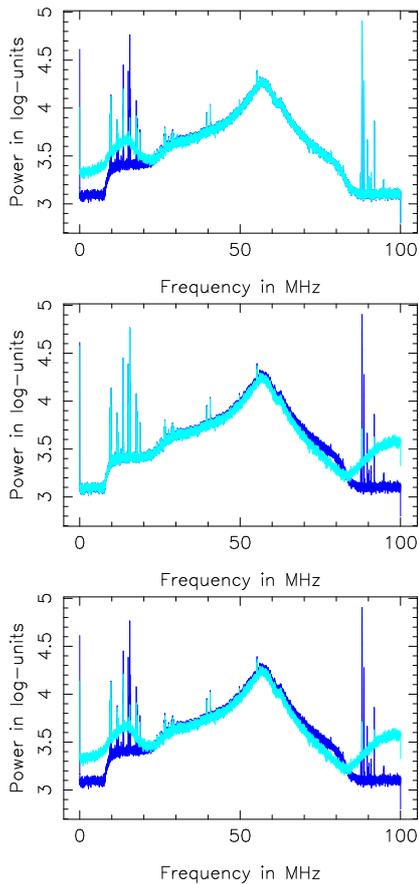

\begin{center}
\includegraphics[angle=-90,width=0.69\linewidth]{eps/TBB-spec-15.eps} 
\includegraphics[angle=-90,width=0.69\linewidth]{eps/TBB-spec-88.eps}
\includegraphics[angle=-90,width=0.69\linewidth]{eps/TBB-spec-1588.eps}
\caption{
 \label{fig:IIR-filters}
Effects of the digital IIR filters on the spectrum: 
unmodified spectrum in dark color and in light color:
top: spectrum with notch-filter at 15\,MHz; 
middle: spectrum with notch-filter at 88\,MHz; 
bottom: spectrum with both filters.
   }
\end{center}
\end{figure}

With IIR filters one can do frequency filtering of a data stream in the time-domain
without a Fourier transform of the data. They work by evaluating:
\begin{eqnarray*}
	x'_i & = & b_0 x_i + b_1 x_{i-1} + b_2 x_{i-2} \dots \\
             &   & - a_1 x'_{i-1} - a_2 x'_{i-2} - a_3 x'_{i-3} \dots
\end{eqnarray*}
where $x'_i$ represents the filtered, output data and $x_i$~is the input data.
They are called {\em infinite impulse response} because in general they are able 
to produce an infinite nonzero output from an input with only one nonzero sample.
In LOFAR we implemented a three stage filter in which the parameters 
$b_0, b_1, b_2, a_1,$ and $a_2$ can be set. By choosing different values for the
parameters one can implement a high-pass, low-pass or notch filter.
One problem when choosing filter parameters is that setting a filter close
to an edge of the frequency band produces aliasing effects, that can distribute
power from one frequency to a previously clean frequency. 
Figure \ref{fig:IIR-filters} shows the effect of two notch filters for the short-wave 
band and the FM transmitter on the frequency spectrum. The aliasing effects are 
clearly visible, but the total RFI power is still reduced by the filters. Test 
of the effect on the signal to noise for the detection of a standard pulse added 
to different noise data showed that the 88\,MHz filter increases the SNR at all 
times, while the 15\,MHz slightly decreases the SNR during times of low RFI from
the short-wave band, but enhances the SNR when significant short-wave band RFI 
is present. While a reduced SNR is not desirable, the increased SNR during times 
of strong RFI evens out changes of the SNR with time of day, which makes also the 
15\,MHz filter desirable.

\section{Particle Detector Array}

\begin{figure}
\begin{center}
\includegraphics[width=0.95\linewidth,clip=true]{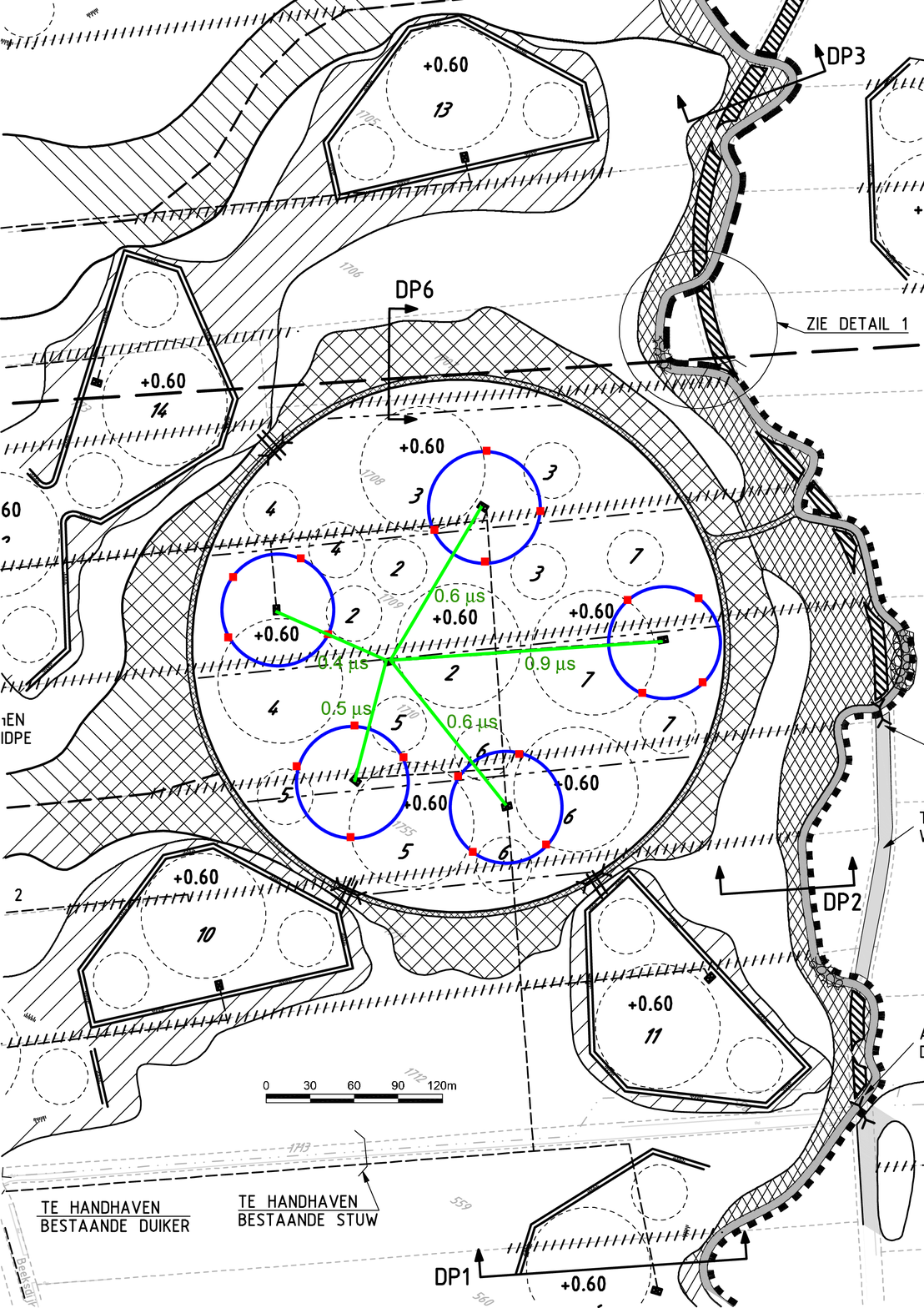}
\caption{
 \label{fig:lasa}
Layout of the LOFAR air shower array (LASA) in the LOFAR core. The small and big 
dashed circles show places of the LOFAR antenna fields. The five solid (blue) 
circles mark 50\,m distance from the electronics stations and the (red) squares on 
these mark the planned positions for the particle detectors.
   }
\end{center}
\end{figure}

In addition to the radio only trigger we will set up a small particle detector 
array in the LOFAR core. The purpose of this array is to test and 
confirm the performance of the radio trigger and provide additional information 
for the air shower analysis. Measuring with both, radio antennas and particle 
detectors will provide an important confirmation of our ability to select real 
air shower events only from radio data.

The array will consist of five stations with four detectors each, placed in the 
so called super-station. Figure~\ref{fig:lasa} shows 
the layout of the array in the LOFAR center.
The electronics of each station will be housed in electronic cabinets of the central 
LOFAR stations. Each station of four detectors will use the electronics developed for 
the HISPARC experiment\cite{hisparc} and have its own electronics with
GPS-timing, trigger logic, and DAQ-PC. Thus it could be run as a stand-alone experiment,
but it can also be triggered from outside, e.g. by the other stations. The data from all 
stations is collected at a central DAQ-PC, which generates combined event-files and can 
send a trigger signal to LOFAR.

The detectors are plastic scintillator detectors previously used in the 
KASCADE\cite{KASCADE} experiment. 
Recent measurements revealed that the detectors do not emit significant RFI and 
the little emission they do emit is not expected to pose a problem for the
regular operation of LOFAR.

\section{Summary}

The high sensitivity and excellent calibration of LOFAR will make it an unique tool
for the measurements of radio emission from air showers. At LOFAR we will measure 
air showers in two modes: In the HECR mode we trigger on radio pulses in already 
beam-formed data, which will limit us to relatively low primary energies. In the VHECR
mode we trigger on pulses in single channel data. This trigger algorithm will have
three steps: on the hardware that writes the raw ADC data into a memory buffer,
on the PC controlling a LOFAR station, and on the central processing facility.
The algorithms for the first step and a preliminary version of the second step have
already been implemented and the first tests have been successful.
We also plan to set up a small particle detector array which will allow us to confirm
that we can pick out real air shower events only with our radio trigger.



\end{document}